\documentclass[aps,twocolumn,showpacs,preprintnumbers,amsmath,amssymb,floatfix]{revtex4}

\usepackage{graphicx}
\usepackage{dcolumn}
\usepackage{bm}
\usepackage[german,american]{babel}

\begin{document}

\title{ Can the dynamics of an
atomic glass-forming system be described as a continuous time random
walk?}

\renewcommand{\thefootnote}{\fnsymbol{footnote}}

\author{Oliver Rubner}
\author{Andreas Heuer}

\affiliation{University of M\"unster, Institute of Physical Chemistry,
Corrensstr. 30, D-48149 M\"unster, Germany}

\begin{abstract}

We show that the dynamics of supercooled liquids, analyzed from
computer simulations of the binary mixture Lennard-Jones system, can
be described in terms of a continuous time random walk (CTRW). The
required discretization comes from mapping the dynamics on
transitions between metabasins.  This comparison involves verifying
the conditions of the CTRW as well as a quantitative test of the
predictions. In particular it is possible to express the wave
vector-dependence of the relaxation time as well as the degree of
non-exponentiality in terms of the first three moments of the
waiting time distribution.

\end{abstract}

\maketitle

The dynamics of supercooled liquids is a very complex process with
many non-trivial features such as non-exponential relaxation,
decoupling of diffusion and relaxation, significant correlated
forward-backward processes (e.g. cage effect), and increasing length
scales of relaxation, just to mention some of the most prominent
\cite{Ediger:427,Debenedetti01,Dyre:2006}. The complexity of the
dynamics originates from the highly cooperative nature of the
dynamical processes.

Several phenomenological models have been proposed which attempt to
describe the dynamics of supercooled liquids in relatively simple
terms, thereby implying some kind of coarse-graining to get rid of
the microscopic details of the dynamics. In the free-energy energy
\cite{Brawer:1984,Dyre:1987,Monthus:310,Diezemann:1998} and the RFOT
models \cite{Xia01} the system relaxes, possibly in a multi-step
process, between different states. One prominent example is the trap
model \cite{Dyre:1987,Monthus:310}, postulating a sequence of escape
processes where the waiting time $\tau$ in a configuration is fully
governed by its energy and the new configuration is randomly chosen
from the set of all possible configurations. Thus, the dynamics is
fully described by the waiting time distribution $\varphi(\tau)$.
Extending this model by the spatial aspects of the relaxation
processes one would, it its simplest version, end up with a
continuous-time random walk (CTRW). Note that in general
continuous-time random walks (CTRWs)
\cite{Montroll:1965,Monthus:310,Odagaki:328,Barkai:2003} as well as
the related Levy walks \cite{Fogedby1,Sokolov} are often used to
describe anomalous dynamic properties, characterized by non-trivial
power-law behavior of quantities such as the mean-square
displacement.

In recent years the facilitated spin models have been revitalized to
grasp the dynamics of supercooled liquids
\cite{Fredrickson84,Garrahan02,Garrahan:2003,Jung2004}. They are
thought to reflect the heterogeneous mobility field of molecular
glass-forming systems. One spin corresponds to a small volume which
is either unjammed or jammed (spin up or down). The ability of a
spin to flip is exclusively governed by the orientation of the
adjacent spins. Self-diffusion has been introduced by postulating a
random walk of the particle with the chance to move if the old as
well as the new site is mobile
\cite{Jung2004,Berthier:2005a,Jung2005}. This dynamics is also
described in terms of a CTRW although for the model variant (East
model), supposed to describe fragile systems, a direct mapping is
not possible \cite{Berthier:2005a}.

Using the CTRW picture in the context of these phenomenological
models does not necessarily imply that it is of relevance for
microscopic glass-forming systems. Here we analyze a binary
Lennard-Jones system (BMLJ), a standard model of supercooled
liquids  \cite{Kob1995}, via computer simulations. The goal of
this work is threefold. First, we explicitly show that the
required conditions for the applicability of the CTRW approach are
fulfilled to a very good approximation. Second, we prove that the
CTRW approach not only allows one to obtain the wave-vector
dependent relaxation time (as already discussed, e.g., in
\cite{Berthier:2005a}) but also the non-exponentiality of
relaxation. Third, the predictions are verified by explicit
comparison with the numerical simulations.

We analyze a BMLJ system with N=65 particles at $T = 0.5$ which is
slightly above the mode-coupling temperature. It has been shown
that this system is large enough to recover the diffusion constant
without significant finite size effect in the range of
temperatures accessible by computer simulations \cite{H9,H10}.
Details of the model are described elsewhere \cite{Kob1995,H9}.
The discretization of the dynamics, required for the application
of the CTRW approach, results from the use of inherent structures,
i.e. local minima of the potential energy landscape
\cite{Stillinger:222,Stillinger:245}, or the use of metabasins
(MBs)\cite{H10}. Of particular relevance in this work is the
incoherent scattering function $S(q,t) = \langle \cos q[x(t + t_0)
- x(t_0)] \rangle$ where the brackets denote the average over all
particles and all $t_0$. Furthermore, $x(t)$ is the x-coordinate
of a particle. The first decay at short times to a value $f < 1$
is due to the fast $\beta$-relaxation whereas the long-time
relaxation reflects the $\alpha$-relaxation. It is often described
by a KWW function $f \exp[-(t/\tau_{KWW})^{\beta_{KWW}}]$. When
analyzing $S(q,t)$ for the sequence of inherent structures rather
than actual configurations it turns out at temperatures close to
the mode-coupling temperature that the short-time decay disappears
and the decay is fully related to the $\alpha$-relaxation with
identical values $\tau_{KWW}$, $\beta_{KWW}$ \cite{Schroder:210}.
Not surprisingly, the same holds for the sequence of MBs (data not
shown). From now on, $S(q,t)$ will represent the case of MBs,
thereby describing the  $\alpha$-relaxation.

Two important observables enter the CTRW approach: (i) the waiting
time distribution $\varphi(\tau)$, (ii) the probability $\pi_1(x)$
that a particle during a transition between two MBs moves a specific
distance along some fixed direction (here: x). More generally,
$\pi_n(x)$ expresses the corresponding probability after $n$ MB
transitions. The Fourier transform is denoted $\pi_n(q)$. Under
conditions (C1)-(C3), which form the basis of the CTRW approach and
are discussed below, it is possible to express $S(q,t)$ in terms of
$\varphi(\tau)$ and $\pi_1(q)$.

(C1) {\it $\pi_1(x)$ does not depend on the waiting time since the
previous transition}. From the data in Fig.1 the validity of (C1)
directly emerges. Only for the longest waiting times, which only
have a very low probability (as reflected by the noise), minor
deviations occur.  As a consequence the spatial and temporal
contributions separate to a very good approximation and one can
write
\begin{equation}
\label{sctrw} S(q,t) = \lim_{N\rightarrow \infty} S_N(q,t) =
\lim_{N\rightarrow \infty} \sum_{n=0}^N S_n(t) \pi_n(q).
\end{equation}
Here $S_n(t)$ denotes the probability to have exactly $n$
transitions during time $t$ . This is the central equation of the
CTRW because it expresses the total dynamics during time $t$ in
terms of discrete processes with well-defined probabilities.

\begin{figure}
\includegraphics[width=8cm]{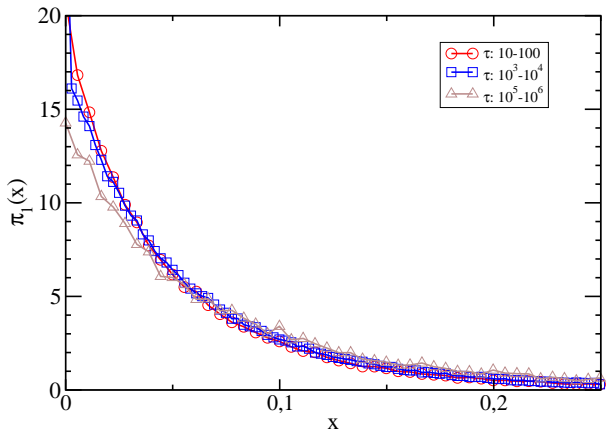}
\caption{\label{fig1} $\pi_1(x)$ for different values of the waiting
times before the corresponding MB transition. Within statistical
uncertainty $\pi_1(x)$ does not depend on the waiting time. }
\end{figure}

(C2) {\it Successive waiting times are statistically uncorrelated so
that the time evolution can be regarded as a sequence of randomly
chosen waiting times.} This has been already shown in
Ref.\cite{H18}. Therefore $S_n(t)$ can be expressed in terms of the
waiting time distribution $\varphi(\tau)$
\cite{Berthier:2005a,review} (see Eq.\ref{sn_lambda} below). Using
the numerically determined $\varphi(\tau)$ and $\pi_n(q)$, one can
compare $S(q_{max},t)$, obtained from simulation, with the
estimation Eq.\ref{sctrw} where $q_{max}$ is the maximum of the
structure factor; see Fig.2. The agreement is very good except for
minor deviations for very long times. Of the order of $10^2$ MB
transition processes are required to have complete relaxation.

\begin{figure}
\includegraphics[width=8cm]{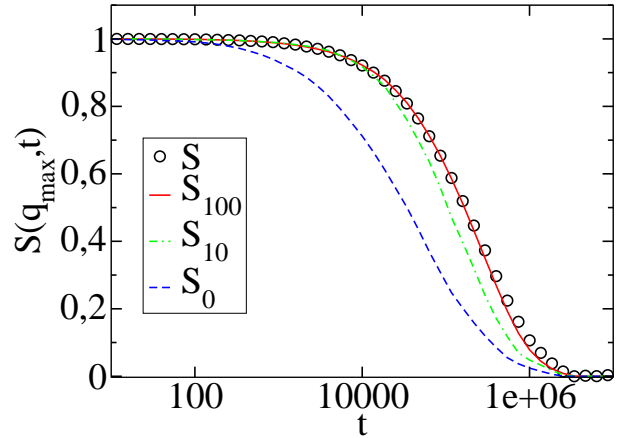}
\caption{\label{fig2} Comparison of  the actual incoherent
scattering function $S(q_{max},t)$ with the estimated function
$S_N(q_{max},t)$ for different values of $N$.}
\end{figure}

(C3) {\it Subsequent transitions are spatially uncorrelated.} The
underlying Markov hypothesis can be formally written as
\begin{equation}
\pi_{n}(x) = \int dx^\prime \pi_{n-1}(x^\prime) \pi_1(x-x^\prime).
\end{equation}
In Fourier-space this convolution reads
\begin{equation}
\label{eq_q} \pi_{n}(q) = \pi_1(q)^n.
\end{equation}
From the analysis of the mean square displacement (MSD) in
Ref.\cite{H9} it became clear that there exist minor
forward-backward correlations for the MB transitions so that (C3)
cannot hold in a strict sense. However, due to the expected locality
of forward-backward transitions one may expect that for longer
length scales, i.e. smaller $q$, they become irrelevant. Indeed, one
has a well-defined limit $a^2 \equiv \lim_{n \rightarrow \infty}
(1/n)\langle x^2 \rangle_{\pi_n} = 0.005$ which is slightly smaller
than $\langle x^2 \rangle_{\pi_1} = 0.009$ \cite{H9} where $\langle
x^2 \rangle_f \equiv \int \, dx \, x^2 f(x)$. To check this in
detail, we have analyzed the n-dependence of $\pi_{n}(q)$, shown in
Fig.3 for different values of the wave-vector $q$. Interestingly,
for $q=q_{max}$ the limiting behavior $ \pi(q)
\equiv(\pi_n(q))^{1/n} = const$ is already reached for $n \ge 5$, as
reflected by the straight line. For smaller $q$-values Eq.\ref{eq_q}
holds even better. Since in the range of relevant $q$ values one has
$a^2 q^2 \ll 1$ the term $\pi(q)$ can be approximated by
$1-q^2a^2/2$. Using inherent structures rather than MBs the large-n
regime would be only reached for $n \approx 10^3$ \cite{H9}. This
would strongly invalidate (C3).

\begin{figure}
\includegraphics[width=8cm]{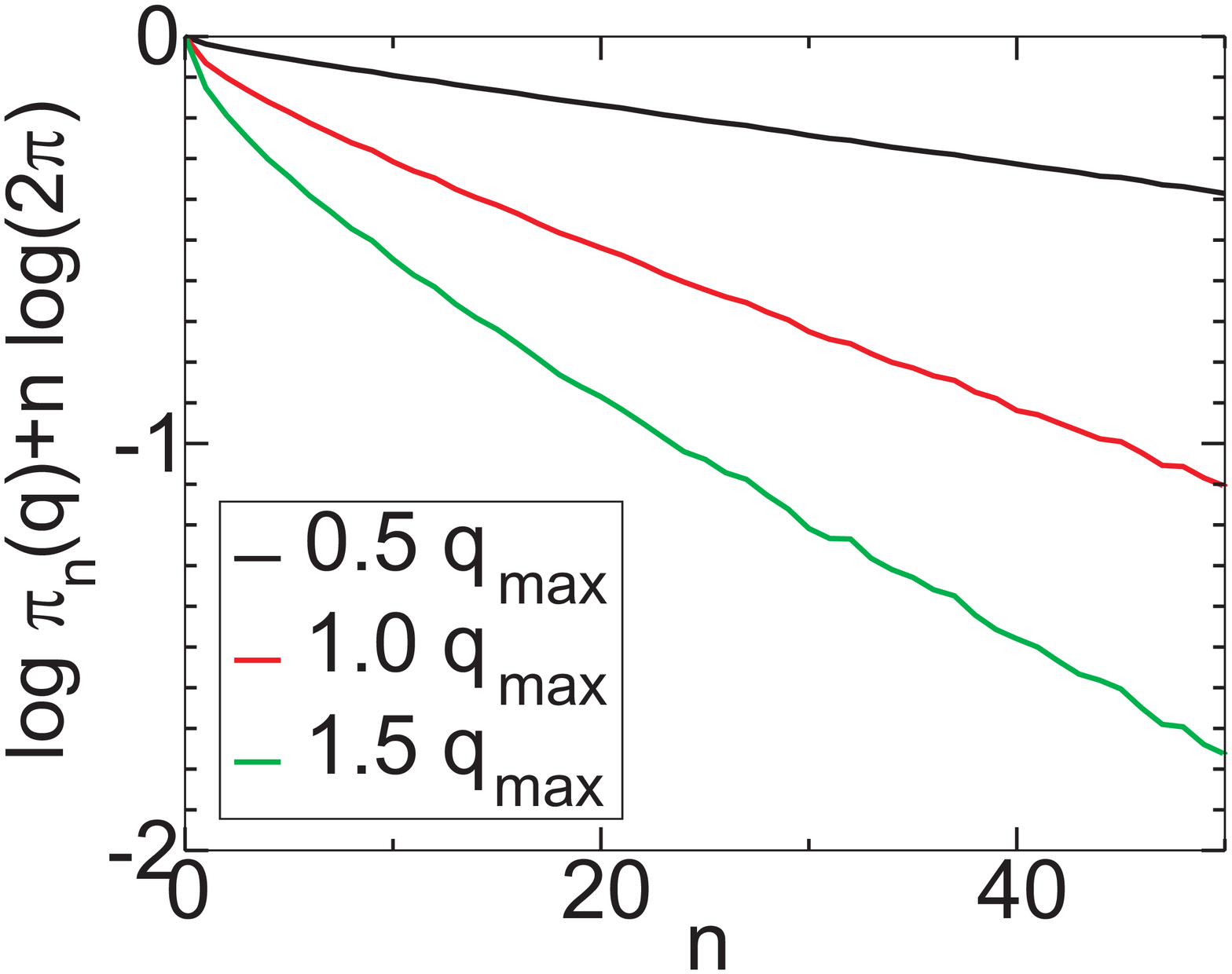}
\caption{\label{fig3} The $n$-dependence of $\pi_n(q)$ for different
values of $q$.}
\end{figure}

Using (C1)-(C3), and substituting all $\pi_n(q)$ by $\pi(q)$, the
temporal Laplace transform of the incoherent scattering function,
i.e. $S(q,\lambda)$, can be calculated analytically, yielding the
Montroll-Weiss equation \cite{Montroll:1965}. Unfortunately, the
inverse Laplace transform of $S(q,\lambda)$ cannot be analytically
performed to calculate $S(q,t)$. Therefore we proceed in a somewhat
different way. First, we define
\begin{equation}
\tau_0(q) \equiv \int dt S(q,t)
\end{equation}
and
\begin{equation}
\label{betadef} {\beta}_m(q) \equiv \frac{\tau_0^2(q)}{\int dt \, t
\, S(q,t)}.
\end{equation}
$\tau_0(q)$ denotes the relaxation time at wave vector $q$ and $
{\beta}_m$ reflects the shape of $S(q,t)$, based on the different
moments. Whereas for exponential relaxation one has $ {\beta}_m =
1$ it decreases when $S(q,t)$ decays in a non-exponential manner.
In case of KWW relaxation one has $ {\beta}_m =
\Gamma^2(1/\beta_{KWW}) / (\beta_{KWW} \Gamma(2/\beta_{KWW}))$
where $\Gamma()$ denotes the $\Gamma$-function (e.g. $\beta_{KWW}
= 1/2$ corresponds to $ {\beta}_m = 1/3$).  $ {\beta}_m$ depends
monotonously on $\beta_{KWW}$ . Thus, $ {\beta}_m$ is a measure of
the degree of non-exponentiality.

Our goal is to find simple expressions for $\tau_0(q)$ and $
{\beta}_m(q)$. For this purpose one can introduce the {\it
persistence time} distribution $\xi(\tau)$. It denotes the
probability that for a random starting point in time the next
transition occurs a time $\tau$ later \cite{Montroll:1965,Jung2005}.
It is related to the waiting time distribution via
\begin{equation}
\xi(\tau) = \int_\tau^\infty dt^\prime \varphi(t^\prime)/\langle
\tau \rangle_\varphi.
\end{equation}
Furthermore, it is related to $S_0(t)$ via
\begin{equation}
S_0(t) = \int_t^\infty \, dt^\prime \xi(t^\prime).
\end{equation}
For $n > 0$ the Laplace-transform of $S_n(t)$ is given by
\begin{equation}
\label{sn_lambda}
S_n(\lambda) = \xi(\lambda)^2
\varphi(\lambda)^{n-1}/\langle \tau \rangle_\varphi.
\end{equation}
Straightforward calculation yields $\int \, dt \, S_n(t) =
S_n(\lambda = 0) = \langle \tau \rangle_\varphi$ for $n > 0$. Note
that for two functions, connected by $f(t) = \int_t^\infty
dt^\prime g(t^\prime)$, one obtains
\begin{equation}
\label{moment} \langle t^n \rangle_f = \frac{\langle
t^{n+1}\rangle_g}{n+1}.
\end{equation}
This implies $\int \, dt \, S_0(t) = \langle \tau \rangle_\xi$, i.e.
the average persistence time. Using again Eq.\ref{moment} it can be
also expressed as $\langle \tau^2\rangle_\varphi/\langle \tau
\rangle_\varphi$. Note that $\langle \tau \rangle_\xi / \langle \tau
\rangle_\varphi \gg 1$ for a broad waiting time distribution,
reflecting large dynamic heterogeneities. Equivalently, this means
that the time until the first transition after a randomly chosen
time takes much longer than the typical time $\langle \tau
\rangle_\varphi$ between successive jumps.

Using Eq.\ref{sctrw} together with Eqs.\ref{eq_q} and \ref{moment}
one obtains \cite{Berthier:2005a}
\begin{equation}
\label{tau0} \tau_0(q)/\langle \tau \rangle_\varphi = \frac{\langle
\tau \rangle_\xi}{\langle \tau \rangle_\varphi} +
\frac{\pi_1(q)}{1-\pi_1(q)} \approx \frac{\langle \tau
\rangle_\xi}{\langle \tau \rangle_\varphi} + \frac{2}{q^2a^2}.
\end{equation}
We note in passing that $\langle \tau \rangle_\xi$ can be identified
with the structural relaxation time $\tau_\alpha$
\cite{Berthier:2005a}. To determine the simulated $\tau_0(q)$ via
integration over $S(q,t)$ we have first fitted $S(q,t)$ by a sum of
two stretched exponentials and then performed the integration
analytically.

In Fig.4a we show the comparison with the simulated data. We have
used $\langle \tau \rangle_\xi/\langle \tau \rangle_\varphi = 27$,
as determined from the numerically determined waiting time
distribution. The $q$-dependence of $\tau_0(q)$ is qualitatively
similar to the data reported in \cite{Berthier2004} and
\cite{Stariolo06}. Note, however, that with the present definition
of $\tau_0(q)$ and the reference to the MB dynamics for the
definition of $\varphi(\tau)$ and $a^2$ a parameter-free prediction
of the $q$-dependence becomes possible. At large $q$ ($q > q_{max}$)
the system relaxes somewhat faster because the effective value of
$a^2$ increases due to the relevance of forward-backward
correlations (see above). For smaller $q$, given by $1/q^2 \approx
\langle \tau \rangle_\xi/\langle \tau \rangle_\varphi a^2/2$, there
is a crossover of $\tau_0(q)$ from the q-independent large-q limit
to the small-q limit $\tau_0(q) = 2 \langle \tau \rangle_\varphi
/(q^2a^2)$. Thus, for large dynamic heterogeneities, i.e. low
temperatures, this crossover may happen at quite large distances
\cite{Berthier:2005a}. Similarly, these non-trivial features are
also reflected by a specific time evolution of the self-part of the
van Hove function $G_s(x,t)$ \cite{Chaudhuri07}. The deviations of
$G_s(x,t)$ from simple diffusion have been analysed in detail in
\cite{Szamel:06}.

\begin{figure}
\includegraphics[width=8cm]{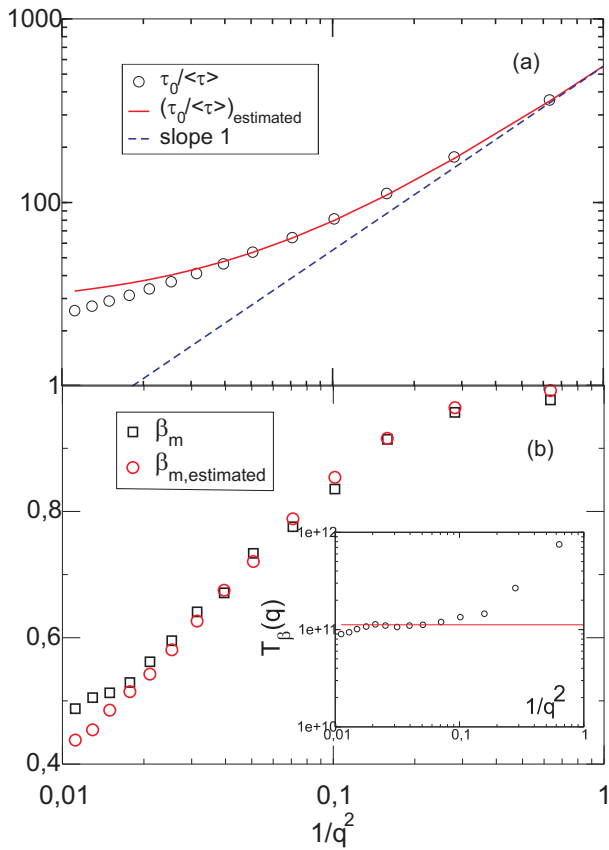}
\caption{\label{t1}  The $q$-dependence of (a) $\tau_0(q)$ together
with its estimation via Eq.\ref{tau0} and (b) of $ {\beta}_m(q)$
together with its estimation. In the inset the validity of the
theoretical expectation $(d/dq) T_\beta(q) = 0$ is tested.}
\end{figure}

For the discussion of $ {\beta}_m(q)$ we first rewrite
Eq.\ref{betadef} as
\begin{equation}
\label{beta_fin}  {\beta}_m^{-1}(q)-1 = T_\beta(q)/\tau_0^2(q)
\end{equation}
with $T_\beta(q) = \int \, dt \, t \, S(q,t) - \tau_0^2(q)$.
Following the standard derivation of the Montroll-Weiss equation one
can show after a tedious but straightforward calculation with the
ingredients, presented in this work, that $(d/dq) T_\beta(q) = 0$
\cite{review}, i.e. $T_\beta(q) \equiv T_\beta$. For the evaluation
of $T_\beta$ we choose the limit $q \rightarrow \infty$ where
$S(q,t) = S_0(t)$. Following Eq.\ref{moment} the first term equals
$\langle \tau^2 \rangle_\xi/2$ whereas the second term is given by
$\langle \tau \rangle_\xi^2$, i.e.
\begin{equation}
\label{T_fin} T_\beta(q) = \frac{\langle \tau^2 \rangle_\xi}{2} -
\langle \tau \rangle_\xi^2
\end{equation}
which directly reflects the width of the persistence time
distribution. Note that via Eq.\ref{moment} $T_\beta$ involves the
third moment of the waiting time distribution $\varphi(\tau)$.
Interestingly, the q-dependence of $ {\beta}_m(q)$ is fully governed
by $\tau_0(q)$. Thus, the degree of non-exponentiality displays
exactly the same crossover-behavior as the relaxation time. A
comparison of Eqs.\ref{beta_fin} and \ref{T_fin} with the numerical
data is shown in Fig.4b, showing again a good agreement. Actually,
due the extreme dependence of the third moment on the fine details
of the long-time behavior of $\varphi(\tau)$ a precise estimation of
$T_\beta$ from $\varphi(\tau)$ is not possible. Again, the
deviations at large $q$ reflect the more complicated dynamics at
short length-scales. The deviations for small $q$ come from the
trivial fact that $T_\beta(q)$ results from a difference of two very
large numbers which, because of the nearly-exponential behavior, are
very similar.

In summary, the present work has shown that the CTRW approach, more
or less explicitly used in different models of the glass transition,
can indeed be {\it numerically derived} for an atomic glass-forming
system. This shows that after an appropriate coarse-graining
procedure (here: the metabasins) the complex dynamics of supercooled
liquids becomes relatively simple. Note that on a lower level of
coarse-graining, namely the inherent structures, (C3) and thus the
CTRW approach is strongly violated. In contrast,  more
coarse-graining, e.g. by joining some successive MBs, would start to
change $S(q,t)$ by rendering it more exponential. In analogy to the
previous model considerations the CTRW approach is formulated for a
subsystem of a large macroscopic system (one cooperatively
rearranging region, one probe molecule, here: a small system with
periodic boundary conditions). Generalization to large systems,
thereby keeping information about possible correlations and
predicting multi-point correlation functions, is a challenge for the
future.

We gratefully acknowledge the support by the DFG via SFB 458 and
helpful discussions with L. Berthier.


\end{document}